# NEW APPROACHES TO INVESTIGATIONS OF THE ANGULAR CORRELATIONS IN NEUTRON DECAY

B. Yerozolimsky

*Harvard University*
*Cambridge, MA 02138*

This article is a restored copy of the text of a talk which was presented on the seminar at NIST in 1996.

ABSTRACT

A new method of measurements of angular correlations in neutron decay is proposed. It excludes the need of precise spectroscopy of decay products and thus promises to make the systematic uncertainties of the results much lower than in experiments carried out up to day.

## 1. INTRODUCTION

The elementary beta-process – the neutron decay has been investigated very intensively during the last thirty years and still remains interesting from the point of view of checking the contemporary theory of Weak Interactions. The absence of the influence of nuclear interaction and the low level of electromagnetic corrections make this process very attractive for this purpose.
As it is well known, the neutron decay leads to the creation of a proton and two leptons – electron and antineutrino, the latter two carrying away the most part of the energy which is released in this process – about 782 KeV.
The probability of neutron decay depends upon the distribution of the energy between the particles and upon the angles between the particle momentums $\vec{P}_e$ and $\vec{P}_\nu$ and the direction of the spin of the decaying neutron $\vec{\sigma}$ in the following way:

$$dW = KG^2 F(E_e) \left[1 + a\frac{\vec{P}_e\vec{P}_\nu}{E_e E_\nu}c^2 + A\frac{\vec{\sigma}\vec{P}_e}{\sigma E_e}c + B\frac{\vec{\sigma}\vec{P}_\nu}{\sigma E_\nu}c + D\frac{\vec{\sigma}[\vec{P}_e\vec{P}_\nu]}{\sigma E_e E_\nu}c^2\right],$$

where $G^2 = g_V^2 + 3g_A^2$, and $g_V$ and $g_A$ are the constants characterizing the intensity of week interaction ( in frame of standard V-A theory).

$$K = \frac{m_e^5 c^4}{2\pi^3 \hbar^7}$$

$F(E_e)$ - is the electron energy spectrum,

$\vec{P}_e, E_e$ and $\vec{P}_\nu, E_\nu$ are corresponding momentums and full energies, c – velocity of light, $\vec{\sigma}$ – the spin of decaying neutron and $a$, $A$, $B$ and $D$ - coefficients of corresponding angular correlations.

The full decay probability $W = 1/\tau_n$ ($\tau_n$ – life time) can be calculated as an integral of $W(E_e, \vec{P}_e, \vec{P}_\nu) dE_e d\Omega_e d\Omega_\nu$

The five values - $\tau_n$, $a$, $A$, $B$, and $D$ can be derived in experiments, and just these five values give the possibility to check the theory predictions.

The contemporary V – A standard theory predicts the following expressions for these parameters characterizing the neutron decay process:

$$\tau_n = \frac{2\pi^3 \hbar^7}{f_n g_V^2 (1+3\lambda^2)}$$

where $\lambda = g_A / g_V$ and $f_n$ – is the phase space factor which can be calculated using the electron spectrum function $F(E_e)$ with taking in account the radiative corrections.

$$a = \frac{1-\lambda^2}{1+3\lambda^2}, \quad A = -2\frac{\lambda+\lambda^2}{1+3\lambda^2}, \quad B = -2\frac{\lambda-\lambda^2}{1+3\lambda^2}, \quad D = \frac{2\,\mathrm{Im}\,\lambda}{1+3\lambda^2}$$

The consideration of pure Fermi type nuclear beta transitions
( $0^+ \to 0^+$ transitions – of such nuclei as $^{14}$C, $^{14}$O, $^{18}$Ne and others)
give for the value of the life time $\tau_{0\to 0}$ the following expression:

$$\tau_{0\to 0} = \frac{4\pi^3 \hbar^7}{f_{0\to 0} g_V^2 m_e^5 c^4} ,$$

where $f_{0\to 0}$ is the integral of the same type as $f_n$ in the case of neutron decay.

The ratio $\dfrac{(f\tau)_{0\to 0}}{(f\tau)_n}$ is equal to $\tfrac{1}{2}(1 + 3\lambda^2)$, and it means that this ratio as well as the values of $a$, $A$, and $B$ are defined by one and the same fundamental parameter of the theory - $\lambda$.

2. THE MODERN* EXPERIMENTAL DATA IN NEUTRON DECAY AND THE STATUS OF V-A THEORY OF WEEK INTERACTION.

*All data which are presented in this paper relate to the experimental situation which was in 1996 .

The modern weighted average values of the constants describe the neutron decay process are the following ones [1], [2] :

| | | | |
|---|---|---|---|
| $\tau_n$ = | 887.0 ±1.6 sec | - | (averaged over 10 last data) |
| $a$ = | - 0.1017 ±0.005 | - | (only two measurements th last one in 1978 ) |
| $A$ = | - 0.1138 ±0.0009 | - | ( averaged over 5 last data) |
| $B$ = | 0.9902 ±0.0080 | - | ( averaged over 3 last data) |
| $D$ = | - 0.0003 ±0.0015 | - | ( two measurements in the middle of 1970 th ) |

The experiments devoted to the measurements of both antineutrino angular correlations are listed in the following table :

| Type of correlation | Year | Laboratory | Result |
|---|---|---|---|
| $\vec{P}_e \vec{P}_\nu$ ($a$ –constant) | 1967 | ITEP (Russia) | -0.091 ±0.039 [3] |
| | 1978 | Zeibersdorf (Austria) | -0.1017±0.0051 [4] |
| $\vec{\sigma} \vec{P}_\nu$ ($B$ –constant) | 1960 | Argonn (USA) | 0.88 ±0.15 [5] |
| | 1961 | Chalk River (Canada) | 0.96 ±0.40 [6] |
| | 1969 | Argonn (USA) | 1,01 ±0.05 [7] |
| | 1970 | Kurchatov Inst. (Russia) | 0.995 ±0.035 [8] |
| | 1995 | PNPI (Russia) Kurchatov Inst. (Russia) NIST (USA) Harvard Univ. (USA) | 0.0894±0.0083 [9] |

These data show how long and difficult was the progress in accuracy of the $B$ – constant, and how low is the accuracy of the constant $a$, which remains to be so poor without any attempts to improve it since the end of 70-th..

The main reason of this situation is connected with the fact that the antineutrino is undetectable, so that the knowledge of its escape direction can be derived only with the help of precise spectrometry of two other particles – electron and the recoil proton. Just this leads to a lot of unpredictable systematic uncertainties and makes the results of measurements unreliable.

The main goal of the proposal which is the subject of this talk is to discuss a new approach to evaluating both antineutrino angular correlations which permits to avoid ( or essentially reduce) these spectrometric difficulties.

But let us look before at the experimental values listed in this section and analyze them from the point of view of modern V – A theory.
Yu.Mostovoy and A.Frank found some simple identities which must be satisfied by experimental data of angular correlation constants $a$, $A$ and $B$ if the V – A theory holds [10].

$$F_1 = 1 + A - B - a \equiv 0$$
$$F_2 = aB - B - A^2 \equiv 0$$

The substitution of experimental data gives:

$$F_1 = -- 0.0023 \pm 0.0095 \quad \text{and}$$
$$F_2 = 0.0002 \pm 0.0052$$

Thus, the experimental data seem to be compatible with theory predictions, and it must be emphasized that the main part of the uncertainty of these calculations is connected with the errors in $a$ and $B$ coefficients.
Including the data of ($f\tau$) values of the neutron decay and of $0^+ \rightarrow 0^+$ transitions in the consideration leads however to a much less degree of compatibility with the theory.
The values of $\lambda$ derived from the $a$ and $A$ experimental data are :

$$|\lambda_a| = 1.259 \pm 0.017$$
$$\lambda_A = -1.2603 \pm 0.0024 ,$$

however , the value of $\lambda$ calculated from ($f\tau$) experimental data with taking in account thoroughly evaluated electromagnetic corrections [11] is:

$$|\lambda_\tau| = -1.2688 \pm 0.0016$$

This discrepancy was recently still larger, and it gave the occasion to some propositions about possible contribution of right-handed V + A current in the week interaction Hamiltonian. But the last experimental data of the $A$ – constant derived by Schreckenbach and others [12] made this discrepancy lower. In any case, the situation needs to be checked very carefully, and new measurements are urgently needed.

The improvement in accuracy of both antineutrino correlation coefficients - $a$ and $B$ – would be very useful from this point of view. All these considerations are illustrated in Fig 1, where the correlation coefficients $a$, $A$ and $B$ are presented as functions of $\lambda$, and all results of calculations of this fundamental ratio from experimental data are shown.

Besides, it is worthwhile to mention that the dependence of $B$ upon $\lambda$ is very slow, thus, the $B$ – value is not suitable for calculating $\lambda$. But at the same time, just this behavior of $B$ makes it very sensitive as a criteria of checking V – A theory. Because if V – A theory is valid, the value of $B$ must be within extremely narrow limits ( $0.988 < B < 0.989$ ) in the whole range of possible values of $\lambda$.

Fig 1 shows that the value of another combination of correlation constants, namely, $A/a$, must also be in a very narrow interval too ( $1.113 < A/a < 1.118$ ).

These considerations confirm the importance of improving the accuracy of experimental data of $B$, $a$ and $A/a$.

3. THE IDEA OF MEASUREMENTS USING TWO DETECTORS WITH SUFFICIENTLY LOW SOLID ANGLES ARRANGED OPPOSITE TO EACH OTHER.

Many years ago a new approach to the problem of investigating the angular correlations involving the antineutrino escape direction which would be free of precise spectrometry problems was the subject of intensive discussions at the Kurchatov institute in Moscow. The participants of those conversations were Yu.Mostovoy, A.Frank and the author of this talk.

The main idea of this approach is quite simple, and it is illustrated in Fig 2. Two detectors ( one for beta electrons and another for recoil protons) are arranged in 180° geometry. They are viewing the source of neutron decay particles - the" decay region" of a polarized neutron beam - in small solid angles defined by appropriate diaphragms.

The energy interval of detected electrons must be far enough from the end of beta –spectrum, and it means that if both solid angles $\Omega_e$ and $\Omega_p$ are small the solid angles of the antineutrino escaping in two opposite directions are small too . This is illustrated in Fig 3 where $\vec{P}_e \vec{P}_v$ and $\vec{P}_p$ are momentums of particles created in the decay process and $\Omega_{v1}$ and $\Omega_{v2}$ are solid angles of antineutrino escape directions when $\vec{P}_v \uparrow\uparrow \vec{P}_e$ and $\vec{P}_v \uparrow\downarrow \vec{P}_e$ respectively. This diagram belongs to a definite electron momentum, but if $\Omega_e$ and the electron energy interval are sufficiently small, the situation remains qualitatively the same. Thus, there are two completely separated groups of recoil protons, corresponding to different antineutrino momentum directions, and these groups can be reliably distinguished with the help of usual time-of –flight technique which does not have to be precise.

Coincidences between the signals of both detectors must be counted in two separated time intervals, which correspond to both antineutrino solid angles.

Unfortunately, these solid angles are different $\Omega_{v1} \neq \Omega_{v2}$, and this is illustrated by the momentum diagram in Fig 3. Thus, it is impossible to derive the value of the *a*-coefficient of electron – antineutrino angular correlation immediately from the counting rates in both groups $N_1$ and $N_2$, because the ratio $\Omega_{v1} / \Omega_{v2}$ must be taken in account, and the calculation of this ratio would need a precise knowledge of the electron energy.

Thus, it was proposed to reverse the polarization of the neutron beam and measure the asymmetries in both groups separately. If the polarization is parallel to the axis of the detector system, the counting rates $N_1^\uparrow$, $N_1^\downarrow$ and $N_2^\uparrow$, $N_2^\downarrow$ in both groups and with opposite directions of the polarization can be written in the following manner:

$$N_1^\uparrow \div (1 + K_{a1} a + K_{A1} A + K_{B1} B) \Omega_e \Omega_{v1}$$
$$N_1^\downarrow \div (1 + K_{a1} a - K_{A1} A - K_{B1} B) \Omega_e \Omega_{v1}$$

and

$$N_2^\uparrow \div (1 - K_{a2} a + K_{A2} A - K_{B2} B) \Omega_e \Omega_{v2}$$
$$N_2^\downarrow \div (1 - K_{a2} a - K_{A2} A + K_{B2} B) \Omega_e \Omega_{v2}$$

where
$K_{a1,2} = <v/c \cdot \cos\theta_{ev}>$
$K_{A1,2} = <v/c \cdot \cos\theta_{e\sigma}> P_n$
$K_{B1,2} = <\cos\theta_{v\sigma}> P_n$

v/c is averaged over the electron energy interval, cosines are averaged over the appropriate solid angles and calculated for the both groups of antineutrino escape directions and $P_n$ is the mean polarization of the neutron beam.

Comparing separately the counting rates belonging to each group we exclude the solid angles $\Omega_e$, $\Omega_{v1}$ and $\Omega_{v2}$ and get the values of the experimental asymmetries $X_{1,2} = \dfrac{N^\uparrow_{1,2} - N^\downarrow_{1,2}}{N^\uparrow_{1,2} + N^\downarrow_{1,2}}$ for both groups :

$$X_1 = \dfrac{K_{A1} A + K_{B1} B}{1 + K_{a1} a} \text{ and } X_2 = \dfrac{K_{A2} A - K_{B2} B}{1 - K_{a2} a}$$

Using these two equations and the value of the $A$ coefficient

(which is known quite well now ) we can calculate the both antineutrino angular correlation constants  $a$ and $B$.
Taking in account the uncertainty in the $A$ – coefficient, the possible accuracy of measurements of the neutron polarization and available intensities of polarized neutron beam we can estimate the final accuracy of the antineutrino angular correlation coefficients, and it turns out that this accuracy in $a$ may be about $\pm 1.5\%$ and in $B$ about $\pm 0.3\%$ after ~ 100 days of measurements.
During the first discussion about this proposal which took place at NIST in August 1994 there was pointed out that the value of $a$ derived in such a measurement can not give an independent way for evaluating the fundamental ratio $\lambda$ of week interaction constants, because it depends very closely upon the value of $A$ substituted. The same comment made Prof. J.Byrne in a letter to the author. And it is real true: a simple analysis shows that the value of $a$ derived from the equations for asymmetries $X_1$ and $X_2$ is almost proportional to the used value of $A$. But it means that as a matter of fact this method can be used for calculating an independent value of the ratio $a/A$ which is also very important for checking the theory predictions as it was discussed earlier and illustrated in Fig 1. The accuracy of the value of $a/A$ derived in such experiment can be estimated as about $\pm 1\%$.

4. THE VARIANT OF MEASUREMENTS WITH THE USE OF A MAGNETIC FIELD LIMITING THE TRANSVERSE COMPONENT OF THE PROTON MOMENTUM.

A new situation arose after Yu.Mostovoy proposed an idea of using a longitudinal magnetic field on the way of recoil protons in order to limit the proton transverse momentum component.
This proposal was described in a preprint of Kurchatov Institute in 1994 [15]. The experimental arrangement is shown in Fig 4.
If the protons make more than one full rotation on their spiral paths towards the detector, the diaphragms which are installed along the axis will stop all of them whose perpendicular component of the momentum $P_{p\perp}$ is more than some $P_{\perp max}$ which depends upon the magnetic field H, diameter of the diaphragm D and upon the distance ρ the point where the proton was created is from the axis.

$$P_{\perp max} \leq \frac{eH}{c}(\frac{D}{2}-\rho)$$

Thus, every point in the decay region of the neutron beam is characterized by a maximal transverse momentum component of the protons which can reach the detector. This confinement does not depend upon the full proton momentum and thus, upon the direction of the antineutrino momentum.

The momentum diagram which illustrates this situation is shown in Fig 5, and, what is most important, the both antineutrino solid angles $\Omega_{v1}$ and $\Omega_{v2}$ are now **identically equal**.

Thus, the value of the antineutrino – electron angular correlation coefficient $a$ can be calculated from a simple formulae :

$$a = \frac{X}{v/c < Cos\theta_{ev} >}$$

where $X = \frac{N_1 - N_2}{N_1 + N_2}$ is the experimental asymmetry of counting rates in both groups of the proton time-of-flight spectrum.

The factors v/c and Cos $\theta_{ev}$ must be calculated, thus, the knowledge of the electron spectrum is needed, but the requirements to the accuracy of spectrometry are not very serious.

The diagram in Fig 5 refers to the case when the electron momentum $\vec{P}_e$ is parallel to the direction of the longitudinal magnetic field H. But if the decay region of the beam and the diaphragm on the way of the electrons are not infinitely small, the electron momentum can be nonparallel to the magnetic field H too.

Fig 6 presents such a situation. The transverse proton momentum component in this case becomes a sum of transverse momentums of both antineutrino and electron : $P_{p\perp} = P_{v\perp} + P_{e\perp}$ .

Besides, the condition of separation of the two recoil proton groups becomes affected too. The cylinder which crosses the sphere of antineutrino momentums and cuts out the regions of $\vec{P}_v$ belonging to the both detected proton groups is now shifted from the central axis , and this causes that the mean values of Cos $\theta_{ev}$ in these groups become different. As the result, the connection between the experimental asymmetry X and the coefficient $a$ we are searching for becomes more complicated. But the solid angles $\Omega_{v1}$ and $\Omega_2$ remain still equal. This is very well seen in Fig 6.

Yu Mostovoy showed that the experimental system discussed may be improved still more, if the recoil protons will be accelerated immediately inside the decay region towards the proton detector with the help of a moderate homogeneous electric field.
 Such an additional field contracts the proton time-of-flight spectrum thus reducing the background of accidental coincidences, and at the same time, it does not spoil the conditions of separation of both proton groups belonging to opposite directions of the antineutrino because the order of arrival of protons with different initial velocities remains unchanged.

But the main advantage which is connected with the use of additional electric field is that protons which initially were moving away from the detector ( $P_v > P_e$ and $\vec{P}_v \uparrow\downarrow \vec{P}_e$ ) are now reflected back towards the detector. As a result, an essentially wider electron energy range can be used which improves the luminosity of the experimental apparatus.

Yu Mostovoy together with S.Balashov carried out computer Monte Carlo calculctions in order to investigate the possibilities of a set-up based on the ideas discussed in [15]. The purpose of these investigations was to check the qualitative picture discussed before and to obtain some preliminary estimations of possible statistical and systematic uncertainties of such a measurement. Proton time-of-flight spectrums were derived with various values of the electron energy intervals and various values of magnetic strength H and electromagnetic field inside the decay region E. One of such spectra belonging to the electron energy range 100 – 300 KeV is shown in Fig 7. The values of the field strengths used were H = 300 Oe , E = 60 V/cm. The response characteristics of beta detector which defines the energy resolution was taken in account too.

The results shown in Fig 7 confirm the main features of the experiment discussed. There are really two well separated groups of protons belonging to decay events with parallel and anti-parallel momentums of the decay leptons. The integral numbers of events in both groups are equal ( the value $a$ =0 was used in these calculations). This fact confirms the main statement about the equality of both antineutrino solid angles.

At the same time, the spectrum in Fig 7 confirm the usefulness of the accelerating field in the decay region : in the most part of electron energy range chosen the initial proton velocity is directed away from the detector , and there would be no protons detected in the second group without this field.

Special tests have been carried out with the response characteristics of the electron detector which takes in account the so called effect of "backscattering "of electrons from the detector which is the source of a long low-amplitude "tail" in the response curve of all solid detectors. Due to the presence of this effect electrons with originally high energies can be mixed up to the chosen energy range of the electron detector , and this might cause the mixing of the groups and make their separation worse.

The result of such test shown in Fig 8 confirms that this effect is really dangerous for the successful realization of the method discussed.

Roughly estimated efficiency of such experimental set-up is about $3 \cdot 10^{-4.}$ . Taking for preliminary estimations the neutron beam

intensity~$10^9$ 1/cm$^2$·sec, the mean neutron velocity ~ $7·10^4$ cm/sec and the volume of the decay region ~50 cm$^3$, one can find that the intensity of the source of neutron decay events will be ~$7·10^2$ 1/sec. Thus, the counting rate of the detector system will take about 0.3 per sec. This means that the accuracy in *a* about 1% may be derived after 100 days of measurements.

In conclusion, it may be important to add that the experimental set-up needed for these series of experiments can be created on the basis of existing apparatus which was built by P.E.Spivack for the purpose of life-time measurements. This apparatus is now at the Kurchatov institute, and a modernization needed for the purpose of carrying out the program discussed is not very complicated and expensive.

It seems, however, that the problems which have to be examined first of all, are the following ones:

1. The problem of backscattering of electrons from the beta detector, which causes a long low amplitude "tail" in the response amplitude distribution.

2. The problems connected with the uniformity of the longitudinal magnetic field ( the end regions, the need of an inlet for the neutron beam etc ).

3. The investigations of the optimal experimental parameters of the detector system.

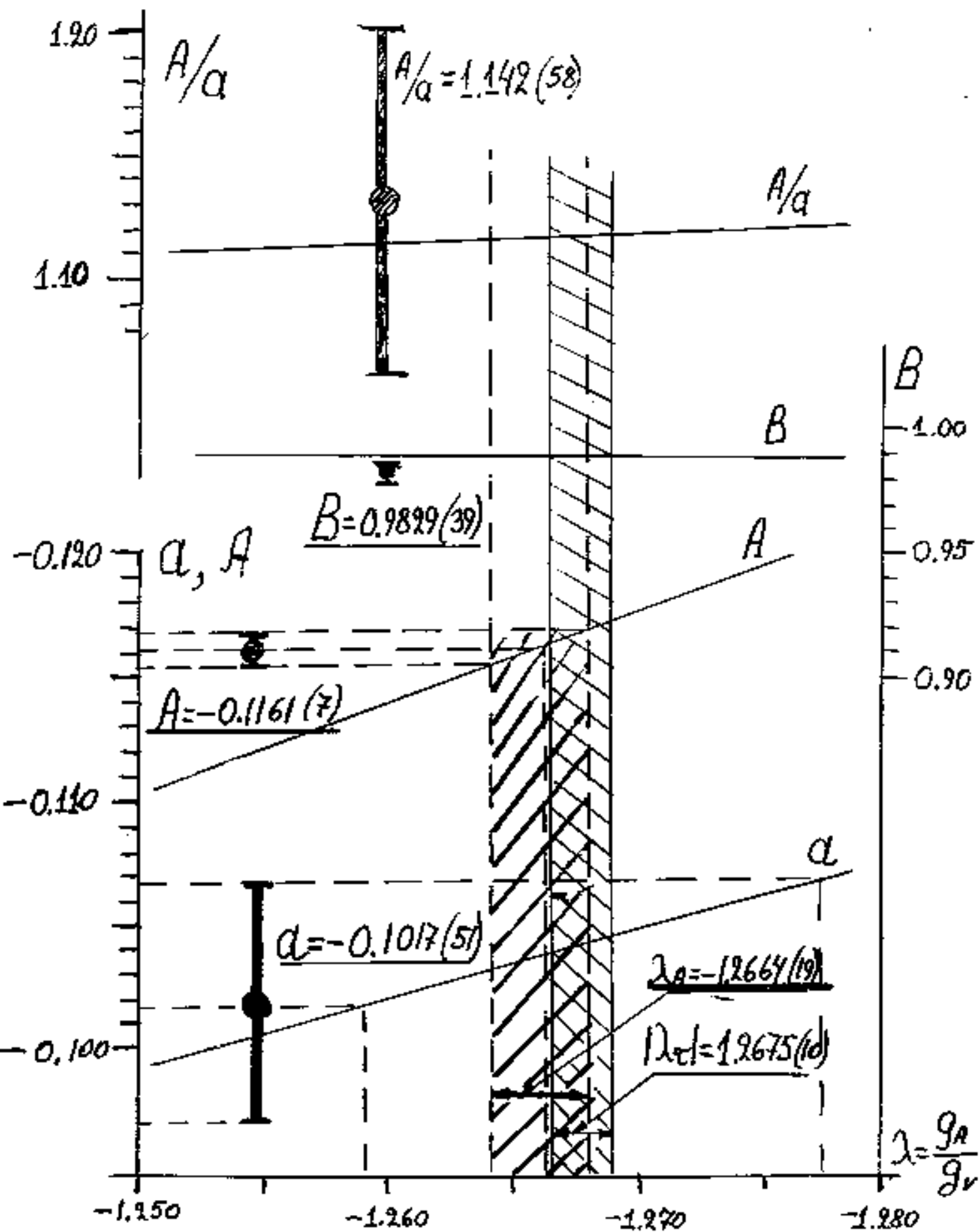

Fig 1

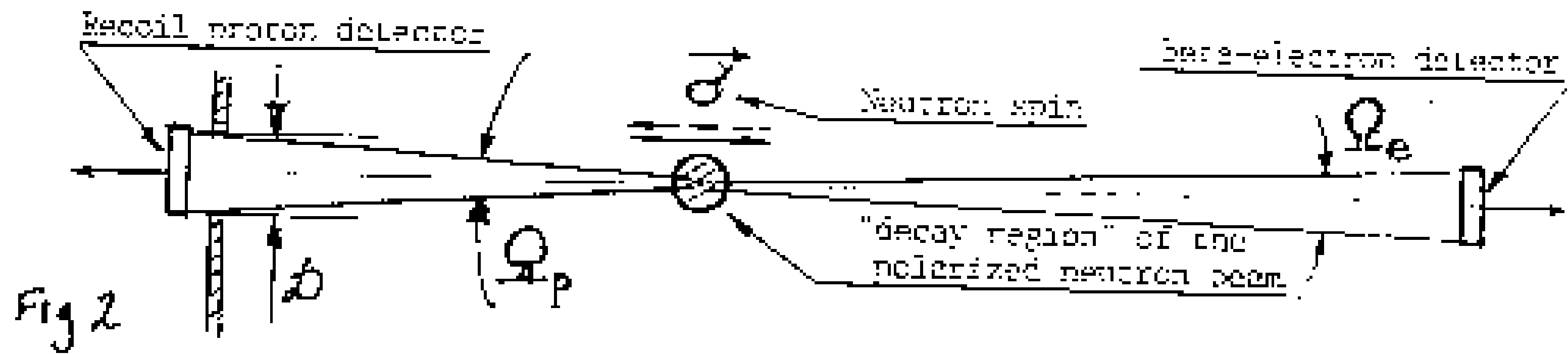

Fig 2

# Momentum diagram

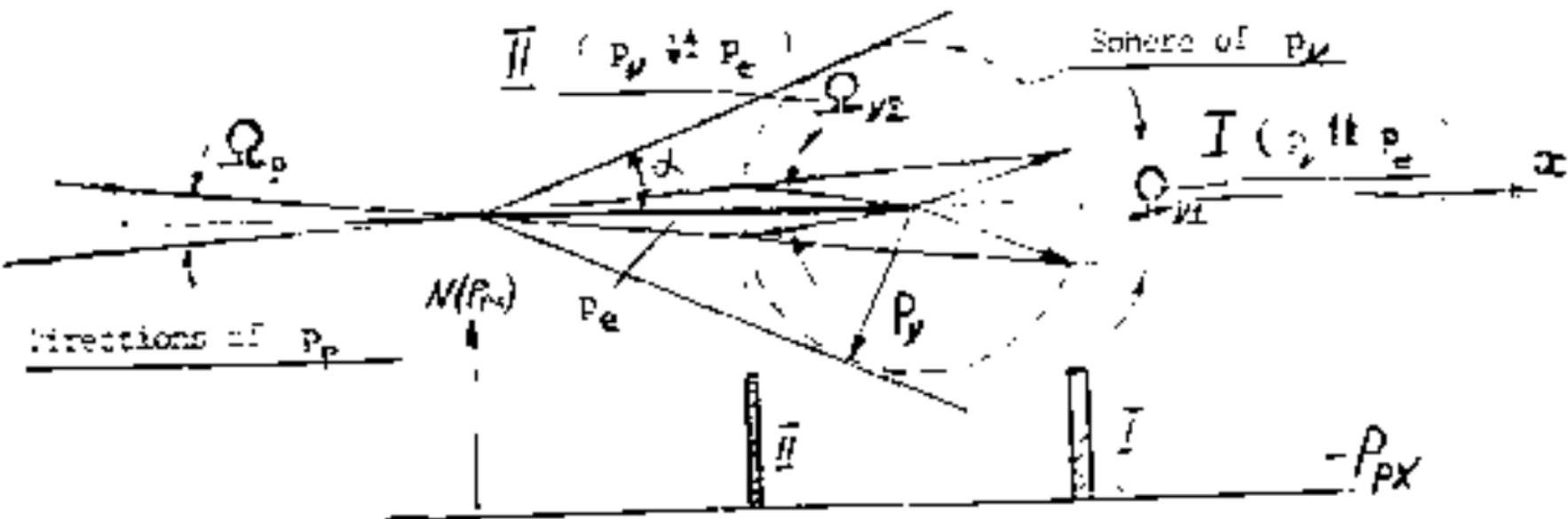

fig 3 — Spectrum of longitudinal component of recoil proton momenta.

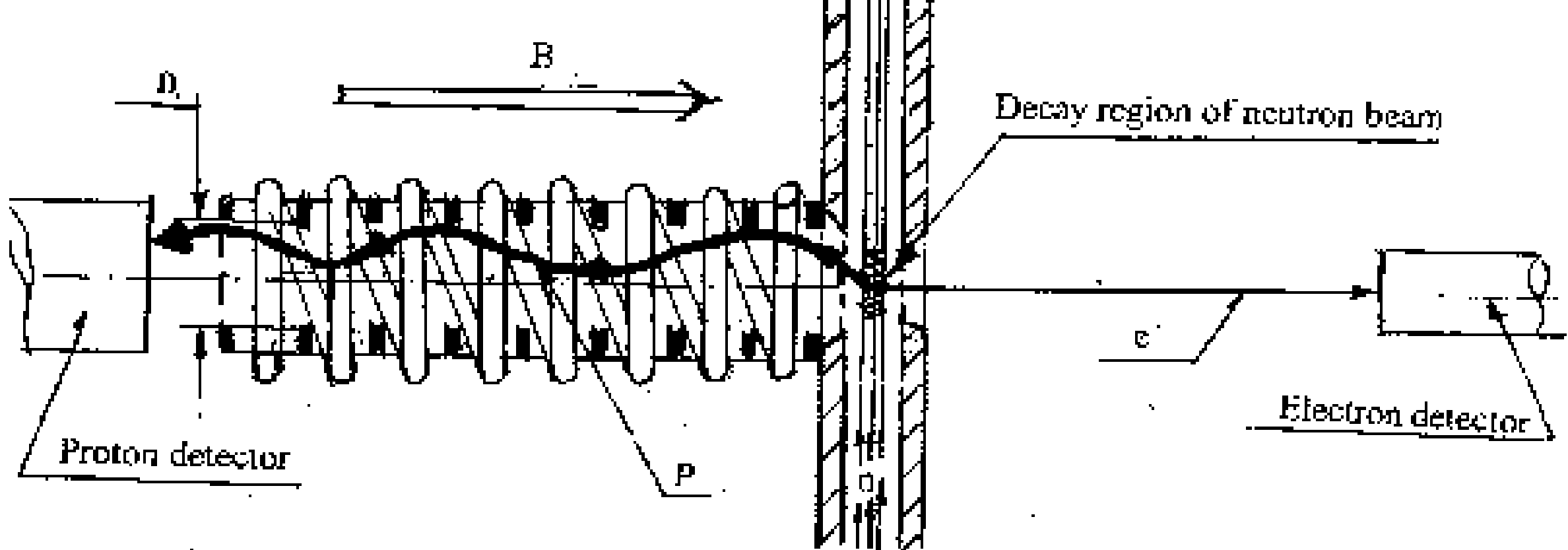

Fig 4  a) The arrangement with the longitudinal magnetic field on the way of recoil protons.

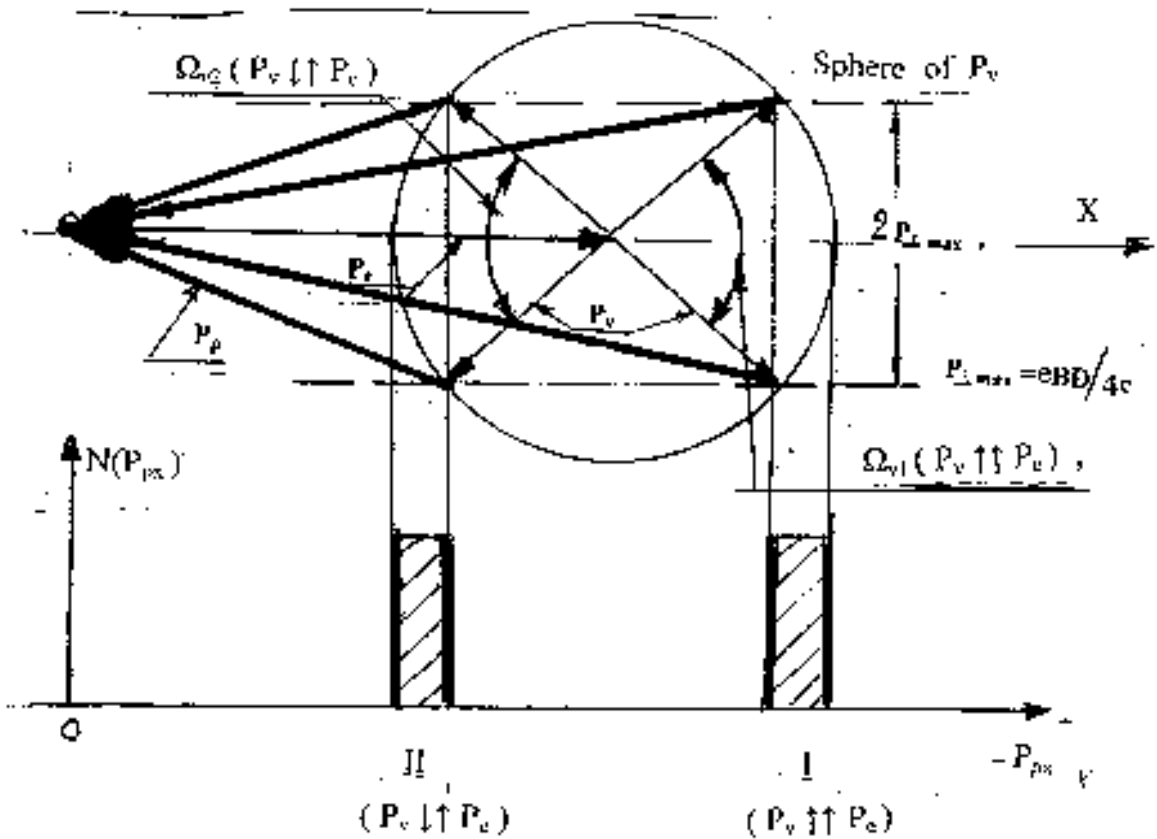

b) Momentum diagram

$\Omega_{v1} = \Omega_{v2}$. This equality remains valid when $P_e \downarrow\uparrow B$ too

Fig. 5

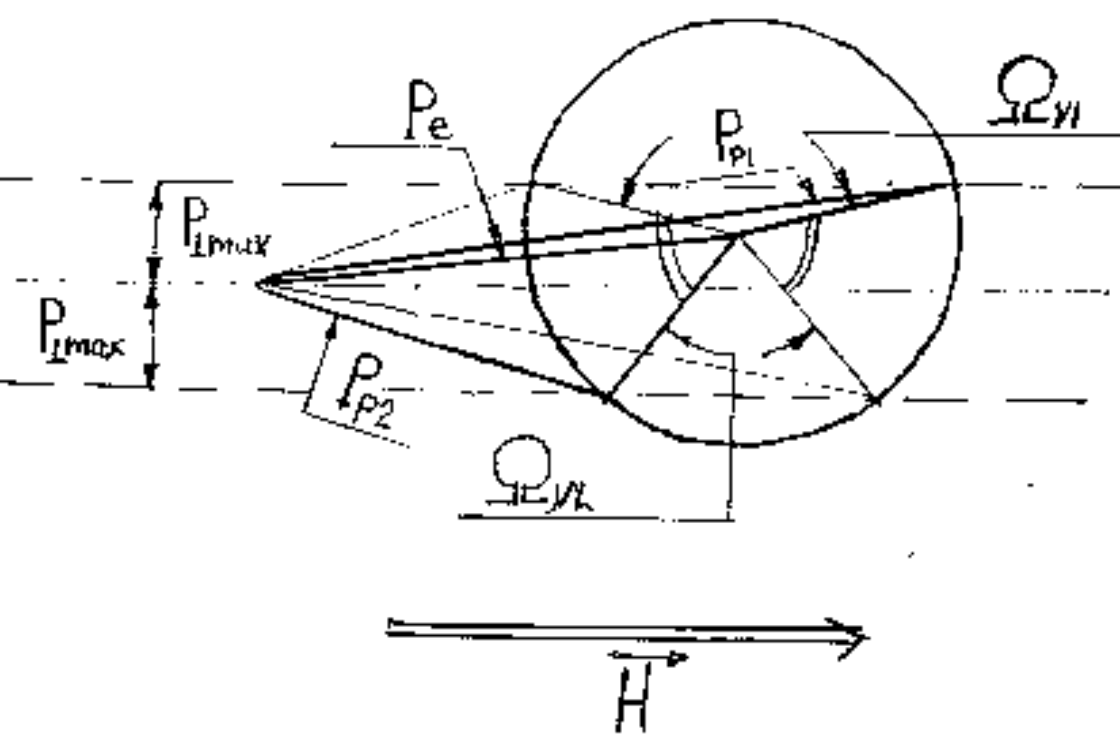

Fig 6

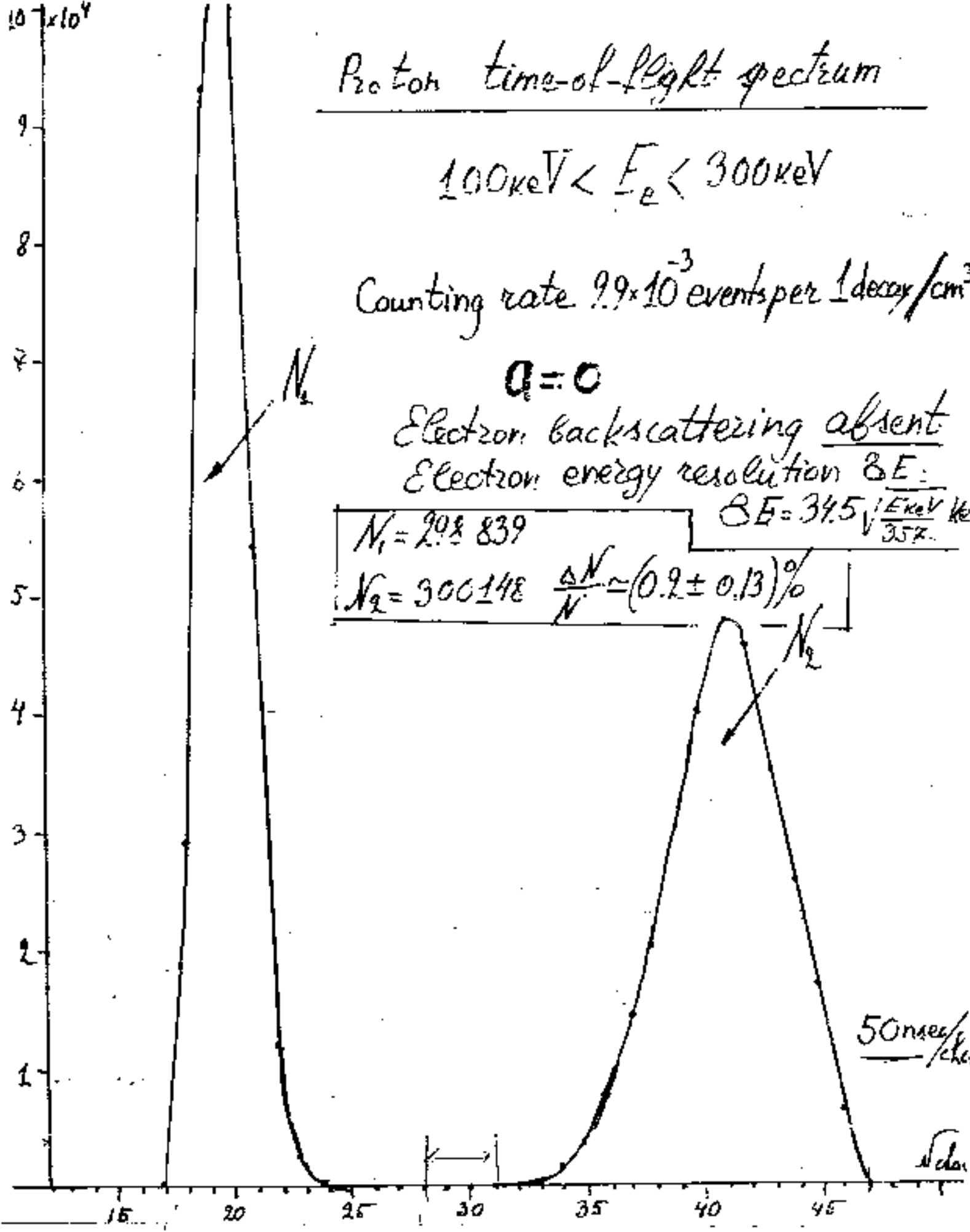

Fig 7

Proton time-of-flight spectrum

$100 keV < E_e < 300 keV$

Counting rate $9.9 \times 10^{-3}$ events per 1 decay/cm$^3$

$a = 0$

Electron backscattering **absent**

Electron energy resolution $\delta E$:

$\delta E = 34.5 \sqrt{\dfrac{E keV}{357}}$ keV

$N_1 = 208\,839$

$N_2 = 300\,148$    $\dfrac{\Delta N}{N} \sim (0.2 \pm 0.13)\%$

50 nsec/ch.

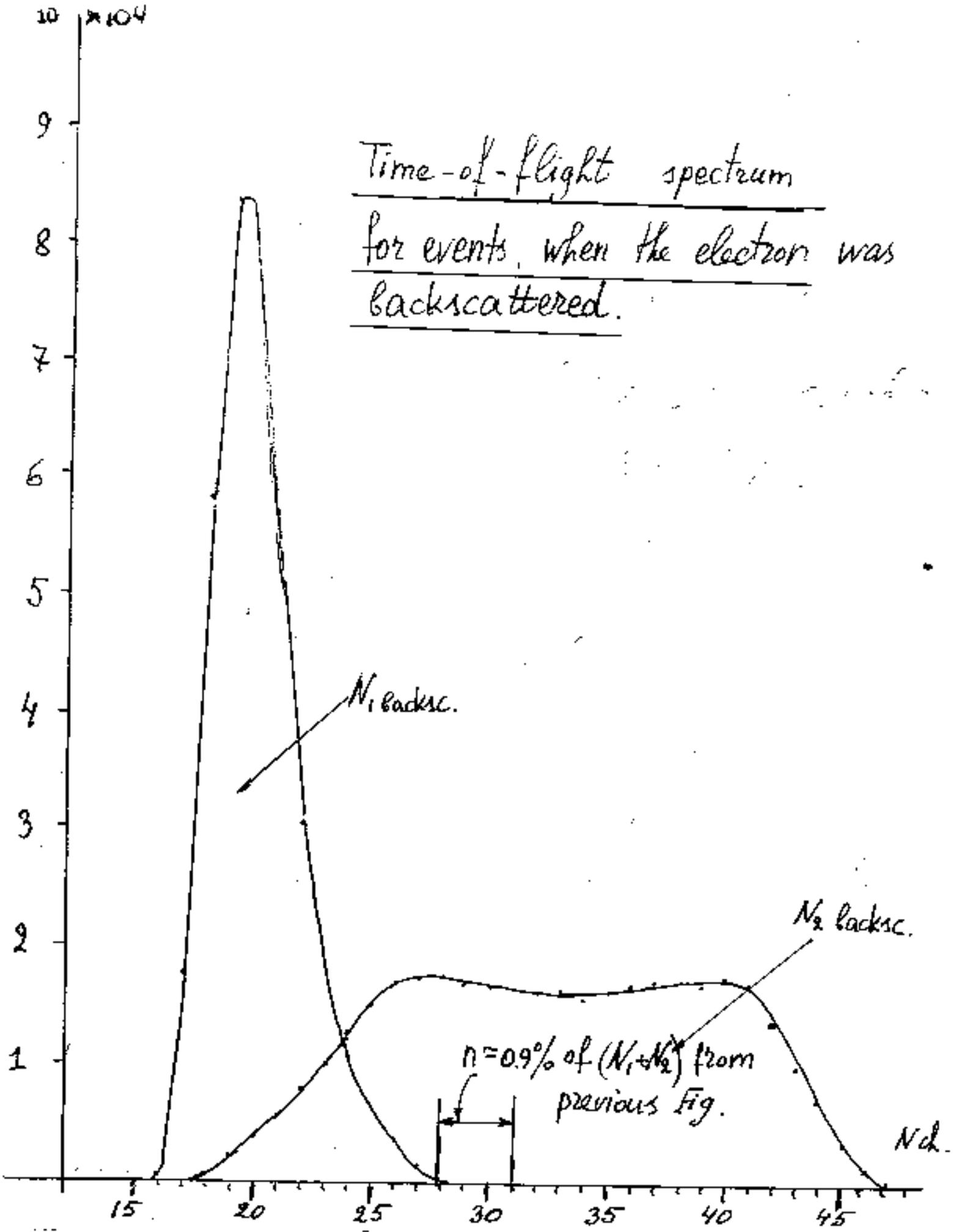

Fig 8